\input{aipcheck}

\documentclass[
    ,final            
  ]
  {aipproc}

\layoutstyle{6x9}

\newcommand{\lsim}{\mathrel{\mathop{\kern 0pt \rlap
  {\raise.2ex\hbox{$<$}}}
  \lower.9ex\hbox{\kern-.190em $\sim$}}}
\newcommand\gsim{\mathrel{\rlap{\lower4pt\hbox{\hskip1pt$\sim$}}
    \raise1pt\hbox{$>$}}}
\def\lsim{\mathrel{\raise.3ex\hbox{$<$\kern-.75em\lower1ex\hbox{$\sim$}}}}
\def\gsim{\mathrel{\raise.3ex\hbox{$>$\kern-.75em\lower1ex\hbox{$\sim$}}}}

\begin{document}

\title{Particle Dark Matter and DAMA/LIBRA}


\author{R. Bernabei}{
  address={Dip. di Fisica, Universit\`a di Roma ``Tor Vergata'', I-00133 Rome, Italy},
  altaddress={INFN, sez. Roma ``Tor Vergata'', I-00133 Rome, Italy}
}
\author{P. Belli}{
  address={INFN, sez. Roma ``Tor Vergata'', I-00133 Rome, Italy}
}
\author{F. Cappella}{
  address={Dip. di Fisica, Universit\`a di Roma ``La Sapienza'', I-00185 Rome, Italy},
  altaddress={INFN, sez. Roma, I-00185 Rome, Italy}
  }
\author{R. Cerulli}{
address={Laboratori Nazionali del Gran Sasso, I.N.F.N., Assergi, Italy}
}
\author{C. J. Dai}{
address={IHEP, Chinese Academy, P.O. Box 918/3, Beijing 100039, China}
}
\author{A. d'Angelo}{
address={Dip. di Fisica, Universit\`a di Roma ``La Sapienza'', I-00185 Rome, Italy},
altaddress={INFN, sez. Roma, I-00185 Rome, Italy}
}
\author{H. L. He}{
address={IHEP, Chinese Academy, P.O. Box 918/3, Beijing 100039, China}
}
\author{A. Incicchitti}{
address={INFN, sez. Roma, I-00185 Rome, Italy}
}
\author{X. H. Ma}{
address={IHEP, Chinese Academy, P.O. Box 918/3, Beijing 100039, China}
}
\author{F. Montecchia}{
address={INFN, sez. Roma ``Tor Vergata'', I-00133 Rome, Italy},
 altaddress={Laboratorio Sperimentale Policentrico di Ingegneria Medica, Universit\`a
di Roma ``Tor Vergata''}
}
\author{F. Nozzoli}{
  address={Dip. di Fisica, Universit\`a di Roma ``Tor Vergata'', I-00133 Rome, Italy},
  altaddress={INFN, sez. Roma ``Tor Vergata'', I-00133 Rome, Italy}
}
\author{D. Prosperi}{
  address={Dip. di Fisica, Universit\`a di Roma ``La Sapienza'', I-00185 Rome, Italy},
  altaddress={INFN, sez. Roma, I-00185 Rome, Italy}
  }
\author{X. D. Sheng}{
address={IHEP, Chinese Academy, P.O. Box 918/3, Beijing 100039, China}
}
\author{R. G. Wang}{
address={IHEP, Chinese Academy, P.O. Box 918/3, Beijing 100039, China}
}
\author{Z. P. Ye}{
  address={IHEP, Chinese Academy, P.O. Box 918/3, Beijing 100039, China},
  ,altaddress={University of Jing Gangshan, Jiangxi, China} 
}

\begin{abstract}
 
The DAMA/LIBRA set-up (about 250 kg highly radiopure
NaI(Tl) sensitive mass) is running at the Gran Sasso National Laboratory
of the I.N.F.N..
The first DAMA/LIBRA results confirm 
the evidence for the presence of a Dark Matter particle component in the galactic
halo, as pointed out by the former DAMA/NaI set-up; cumulatively the data 
support such evidence at 8.2 $\sigma$ C.L. and 
satisfy all the many peculiarities of the
Dark Matter annual modulation signature. The main aspects and prospects 
of this model independent
experimental approach will be outlined.
\end{abstract}

\maketitle


\section{INTRODUCTION}

 DAMA is an observatory for rare processes located deep 
underground at the Gran Sasso National Laboratory of the I.N.F.N.. 
It is based on the development and use of low
background scintillators. The main expe\-ri\-mental set-ups are:
i) DAMA/NaI ($\simeq$ 100 kg of highly radiopure NaI(Tl)) which completed
its data taking on July 2002 \cite{Nim98,allDM,Sist,RNC,ijmd,ijma,epj06,ijma07,chan,wimpele,ldm,allRare};
ii) DAMA/LXe ($\simeq$ 6.5 kg liquid Kr-free Xenon enriched either in $^{129}$Xe or
in $^{136}$Xe) \cite{DAMALXe};
iii) DAMA/R\&D, a facility dedicated to tests on prototypes and to experiments for investigations on rare events \cite{DAMARD};
iv) DAMA/Ge, dedicated to sample measurements and to specific measurements on rare events  \cite{DAMAGE};
v) the second generation DAMA/LIBRA set-up ($\simeq$ 250 kg
highly radiopure NaI(Tl)) \cite{perflibra,modlibra,papep}.
Profiting of the low background features of these set-ups, many rare
processes are studied.

In particular, DAMA/LIBRA is investigating the 
presence of Dark Matter (DM)  particles in the galactic halo 
by exploiting the model independent DM annual modulation signature.
 This signature --
originally suggested in the middle of 80's \cite{Freese} --
exploits the effect of the Earth revolution around the Sun on the number of events induced by
the Dark Matter particles in a suitable low-background set-up placed deep underground.
In fact, as a consequence of its
annual revolution, the Earth should be
crossed by a larger flux of Dark Matter particles around $\sim$ 2 June
(when its rotational velocity is summed
to the one of the solar system with respect to the Galaxy)
and by a smaller one around $\sim$ 2 December (when the two velocities are subtracted).
This offers an efficient model independent signature, able to test a large 
number of DM candidates, a large interval of
cross sections and of halo densities.
It should be stressed that the DM annual modulation is not
-- as often naively said -- a ``seasonal'' variation and it
is not a ``winter-summer'' effect;
in fact, the DM annual modulation is related to the Earth velocity in the galactic frame and
its phase (roughly 2$^{nd}$ June) is well different than those
of physical quantities (such as temperature of atmosphere,
pressure, other meteorological
parameters, cosmic rays flux, ...) correlated with seasons instead.

The DM annual modulation signature is very distinctive
since the corresponding signal must simultaneously satisfy
all the following requirements: the rate must contain a component
modulated according to a cosine function (1) with one year period (2)
and a phase that peaks roughly around $\simeq$ 2$^{nd}$ June (3);
this modulation must only be found
in a well-defined low energy range, where DM particle induced events
can be present  (4); it must apply only to those events in
which just one detector of many actually ``fires'' ({\it single-hit events}), since
the DM particle multi-interaction probability is negligible (5); the modulation
amplitude in the region of maximal sensitivity must be $\lsim$7$\%$
for usually adopted halo distributions (6), but it can
be larger in case of some possible scenarios such as e.g. those in refs. \cite{Wei01,Fre04}.
Only systematic effects or side reactions able to fulfil these requirements and to account
for the whole observed modulation amplitude
could mimic this signature;
thus, no other effect investigated so far in the field of rare processes offers
a so stringent and unambiguous signature.

At present status of technology it is the only model independent signature available in direct dark matter
investigation that can be effectively exploited.

It is worth noting that the corollary questions related to the exact nature of the DM
particle(s) (detected
by means of the DM annual modulation signature)
and to the astrophysical, nuclear and particle Physics scenarios
require subsequent model dependent corollary analyses,
as those performed e.g. in refs. \cite{RNC,ijmd,ijma,epj06,ijma07,chan,wimpele,ldm}.
On the other hand, one should stress that it does not exist any approach
in direct and indirect DM
searches which can offer information on the nature of the candidate in a model independent way, 
that is without assuming some astrophysical, nuclear and particle Physics scenarios.

In the following, we will just briefly summarize the first results
on the Dark Matter particle investigation obtained by DAMA/LIBRA,
exploiting over four annual
cycles the model independent DM annual modulation signature (exposure
of 0.53 ton$\times$yr). The results have also been combined
together with the previous data collected over 7 annual cycles by DAMA/NaI (0.29 ton$\times$yr).
Thus, the whole available data correspond to 11 annual cycles for a total exposure of
0.82 ton$\times$yr, which is orders of magnitude larger than the exposures typically collected in the 
field.

\section{DAMA/LIBRA RESULTS}

The DAMA/NaI set up and its performances are reported in ref.\cite{Nim98,Sist,RNC,ijmd}, while
the DAMA/LIBRA set-up and its performances are described in ref. \cite{perflibra}.
Here we just summarized the main features: i) the sensitive part of the set-up is made of 25
highly radiopure NaI(Tl) crystal scintillators placed in a 5-rows by 5-columns matrix; 
ii) the detectors' responses range 
from 5.5 to 7.5 photoelectrons/keV; iii) the hardware threshold of each PMT is at single 
photoelectron (each detector is equipped with two low background photomultipliers working in 
coincidence); 
iv) energy calibration with X-rays/$\gamma$ sources are regularly carried out down to few keV
in the same conditions as the production runs; 
v) the software energy threshold of the experiment is 2 keV.

\begin{figure}[!ht]
\includegraphics[width=14.5cm]{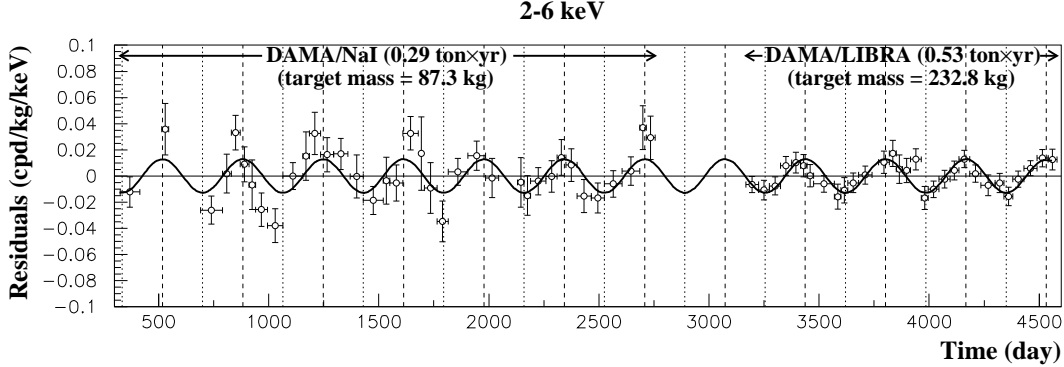}
\vspace{-0.3cm}
\caption{Experimental model-independent residual rate of the {\it single-hit} scintillation events, 
measured by DAMA/NaI and DAMA/LIBRA in the (2 -- 6) keV energy interval 
as a function of the time. 
The zero of the time scale is January 1$^{st}$ 
of the first year of data taking of DAMA/NaI. 
The experimental points present the errors as vertical bars and the associated time bin width as horizontal bars. 
The superimposed curve is the cosinusoidal functions behaviors $A \cos \omega(t-t_0)$  
with a period $T = \frac{2\pi}{\omega} =  1$ yr, with a phase $t_0 = 152.5$ day (June 2$^{nd}$) and with
modulation amplitude, $A$, equal to the central value obtained by best fit over the whole data, that is: 
$(0.0129 \pm 0.0016)$ cpd/kg/keV.
The dashed vertical lines correspond to the maximum of the signal (June 2$^{nd}$), while 
the dotted vertical lines correspond to the minimum.
The total exposure is 0.82 ton $\times$ yr. For details and for other energy intervals 
see \protect\cite{modlibra}.}
\label{fig1}
\end{figure}

Several analyses on the model-independent DM annual 
modulation signature have been performed (see ref. \cite{modlibra} and references therein); 
here just few arguments are mentioned.  
In particular, Fig. \ref{fig1} shows the time behaviour of the experimental 
residual rates of the {\it single-hit} 
events collected by DAMA/NaI and by DAMA/LIBRA 
in the (2--6) keV energy interval \cite{modlibra}.
The superimposed curve in Fig. \ref{fig1} 
represents the cosinusoidal functions behavior $A \cos \omega(t-t_0)$  
with a period $T = \frac{2\pi}{\omega} =  1$ yr and with a phase $t_0 = 152.5$ day (June 2$^{nd}$), 
while the modulation amplitude, $A$, has been obtained by best fit over the DAMA/NaI 
and DAMA/LIBRA data.
When the period and the phase parameters are also released in the fit, values well compatible
with those expected for a DM particle induced effect are obtained \cite{modlibra}:
$T = (0.998 \pm 0.003)$ yr and $t_0 = (144 \pm 8)$ day in the cumulative (2--6) keV energy interval.
Summarizing, the analysis of the {\it single-hit} residual rate favours the presence of a 
modulated cosine-like behaviour with proper features at 8.2 $\sigma$ C.L.  \cite{modlibra}.

The same data of Fig.\ref{fig1} have also been investigated by a Fourier analysis, obtaining 
a clear peak corresponding to a period of 1 year \cite{modlibra}; 
this analysis in other energy region shows instead only aliasing peaks.
Moreover, while in the (2--6) keV {\it single-hit} residuals
a clear modulation is present, 
it is absent at energies just above \cite{modlibra}. In particular, in order 
to verify absence of annual modulation in other energy regions and, thus,  
to also verify the absence of any significant background modulation, 
the energy distribution measured during the data taking periods
in energy regions not of interest for DM detection 
has also been investigated.
In fact, the background in the lowest energy region is
essentially due to ``Compton'' electrons, X-rays and/or Auger
electrons, muon induced events, etc., which are strictly correlated
with the events in the higher energy part of the spectrum;
thus, if a modulation detected 
in the lowest energy region would be due to
a modulation of the background (rather than to a signal),
an equal or larger modulation in the higher energy regions should be present.
The data analyses have allowed to exclude the presence of a background
modulation in the whole energy spectrum at a level much
lower than the effect found in the lowest energy region for the {\it single-hit} events
\cite{modlibra}. 

\begin{figure}[!ht]
\includegraphics[width=13.cm] {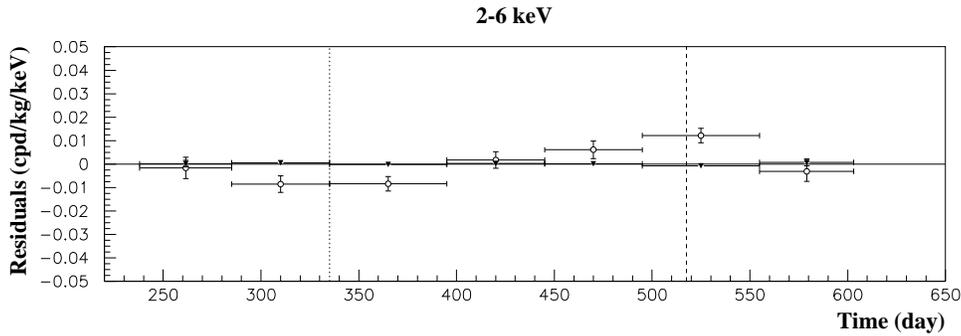}
\caption{Experimental residual rates over the four DAMA/LIBRA annual cycles for {\it single-hit} events (open 
circles) (class of events to which DM events belong) and for {\it multiple-hits} events (filled triangles)
(class of events to which DM events do not belong),
in the energy interval (2 -- 6) keV.
They have been obtained by considering for each class of events the data as collected in a single annual cycle 
and by using in both cases the same identical hardware and the same identical software procedures.
The initial time of the scale is taken on August 7$^{th}$.
The experimental points present the errors as vertical bars and the associated time bin width as horizontal 
bars. See ref. \cite{modlibra}. Analogous results were obtained for the DAMA/NaI data \cite{ijmd}.}
\label{fig_mul}
\end{figure}

A further relevant investigation has been done by applying the same hardware and software 
procedures, 
used to acquire and to analyse the {\it single-hit} residual rate, to the {\it multiple-hits} one. 
In fact, since the probability that a DM particle interacts in more than one detector 
is negligible, a DM signal can be present just in the {\it single-hit} residual rate.
Thus, this allows the test of the background behaviour in the same energy interval of the observed 
positive effect. In particular, Fig. \ref{fig_mul} shows the residual rates of the {\it single-hit} 
events measured over the four 
DAMA/LIBRA annual cycles, as collected in a single annual cycle, together with the residual rates of the {\it 
multiple-hits} events, in the same considered energy interval. 
A clear modulation is present in the {\it single-hit} events, while
the fitted modulation amplitudes for the {\it multiple-hits} 
residual rate are well compatible with zero \cite{modlibra}.
Similar results were previously obtained also for the DAMA/NaI case \cite{ijmd}.
Thus, again evidence of annual modulation with proper features, as required by 
the DM annual modulation signature, is 
present in the {\it single-hit} residuals (events class to which the 
DM particle induced events belong), while it is absent in the {\it multiple-hits} residual rate (event class to 
which only background events belong).
Since the same identical hardware and the same identical software procedures have been used to analyse the 
two classes of events, the obtained result offers an additional strong support for the presence of a DM 
particle component in the galactic halo further excluding any side effect either from hardware or from software 
procedures or from background.

The annual modulation present at low energy has also been analyzed 
by depicting the differential modulation amplitudes, 
$S_{m,k}$, as a function of the energy (the $k$ index identifies the energy interval); the $S_{m,k}$ is the
modulation amplitude of the modulated part of the signal obtained
by maximum likelihood method over the data, considering $T=1$ yr and $t_0=152.5$ day.
The $S_{m,k}$ values are reported as function of the energy in ref. \cite{modlibra}.
It can be inferred that a positive signal is present in the (2--6) keV energy interval, while $S_{m,k}$
values compatible with zero are present just above; in particular, the $S_{m,k}$ values
in the (6--20) keV energy interval have random fluctuations around zero with
$\chi^2$ equal to 24.4 for 28 degrees of freedom.
It has been also verified that the measured modulation amplitudes are statistically well
distributed in all the crystals, in all the annual cycles and energy bins;
these and other discussions can be found in ref. \cite{modlibra}.
It is also interesting the results of the analysis performed by releasing the 
assumption of a phase $t_0=152.5$ day in the procedure of maximum likelihood to 
evaluate the modulation amplitudes from the data of the
seven annual cycles of DAMA/NaI and the four annual cycles of DAMA/LIBRA. In this case
alternatively the signal has been written as:
$S_{0,k} + S_{m,k} \cos\omega(t-t_0) + Z_{m,k} \sin\omega(t-t_0) = 
 S_{0,k} + Y_{m,k} \cos\omega(t-t^*)$, where $S_{0,k}$ is the constant part of the signal in $k$-th energy interval.
Obviously, for signals induced by DM particles one would expect: 
i) $Z_{m,k} \sim 0$ (because of the orthogonality between the cosine and the sine functions); 
ii) $S_{m,k} \simeq Y_{m,k}$; iii) $t^* \simeq t_0=152.5$ day. 
In fact, these conditions hold for most of the dark halo models; however, it is worth noting that 
slight differences in the phase can be expected in case of possible contributions
from non-thermalized DM components, such as e.g. the SagDEG stream \cite{epj06} 
and the caustics \cite{caus}.
The $2\sigma$ contours in the plane $(S_m , Z_m)$ 
for the (2--6) keV and (6--14) keV energy intervals and 
those in the plane $(Y_m , t^*)$ are reported in ref. \cite{modlibra}.
The best fit values for the (2--6) keV energy interval are ($1\sigma$ errors): 
$S_m= (0.0122 \pm 0.0016)$ cpd/kg/keV; 
$Z_m=-(0.0019 \pm 0.0017)$ cpd/kg/keV; 
$Y_m= (0.0123 \pm 0.0016)$ cpd/kg/keV; 
$t^*= (144.0  \pm 7.5)$ day;
while for the (6--14) keV energy interval are:
$S_m= (0.0005 \pm 0.0010)$ cpd/kg/keV;
$Z_m= (0.0011 \pm 0.0012)$ cpd/kg/keV;
$Y_m= (0.0012 \pm 0.0011)$ cpd/kg/keV
and $t^*$ obviously not determined. 
These results confirm those achieved by other kinds of analyses.
In particular, a modulation amplitude is present in the lower energy intervals and the period
and the phase agree with those expected for DM induced signals.
For more detailed discussions see ref. \cite{modlibra}

Both the data of DAMA/LIBRA and of DAMA/NaI
fulfil all the requirements of the DM annual modulation signature. 

As previously done for DAMA/NaI \cite{RNC,ijmd}, careful investigations
on absence of any significant systematics or side reaction effect in DAMA/LIBRA
have been quantitatively carried out and reported in details in ref. 
\cite{modlibra}.
In order to continuously monitor the running conditions, several pieces of information 
are acquired with the production data 
and quantitatively analyzed.
No modulation has been found in any  
possible source of systematics or side reactions for DAMA/LIBRA as well; thus, cautious upper limits 
(90\% C.L.) on the possible contributions to the DAMA/LIBRA measured modulation amplitude
have been estimated \cite{modlibra}.
No systematics or side reactions able to mimic the signature (that is, able to
account for the measured modulation amplitude and simultaneously satisfy 
all the requirements of the signature) has been found or suggested 
by anyone over more than a decade.
For detailed quantitative discussions on all the related topics and for results see ref. 
\cite{modlibra} and refs. therein. 
Just as an example we recall here the case of muons, whose flux has been reported
by the MACRO experiment to have a 2\% modulation with phase around mid--July \cite{Mac97}. In particular, 
it has been shown
that not only this effect would give rise in the DAMA set-ups to a quantitatively
negligible contribution \cite{modlibra,RNC,ijmd}, 
but several of the six requirements necessary to mimic
the annual modulation signature -- namely e.g. the conditions of presence of modulation 
just in the {\it single-hit}
event rate at low energy and of the phase value -- would also fail.
Moreover, even the pessimistic assumption of whatever hypothetical 
(even exotic) possible cosmogenic
product -- whose decay or de-excitation or whatever else 
might produce: i) only events at low energy; ii) only {\it single-hit}
events; iii) no sizeable effect in the {\it multiple-hits} counting rate --
cannot give rise to any side process able to mimic the investigated DM signature.
In fact, not only this latter hypothetical process
would be quantitatively negligible \cite{modlibra}, but 
in addition its phase -- as it can be easily derived -- 
would be (much) larger than July 15th, and therefore well
different from the one measured by the DAMA experiments and expected
by the DM annual modulation
signature ($\simeq$ June 2nd). 
Recently, a LVD analysis \cite{LVD} has been reported for the muon flux relatively to the period 2001--2008, which partially
overlaps the DAMA/NaI running periods and completely those of DAMA/LIBRA.
A value of $\simeq 185$ days has been measured by LVD in this period for the muon phase to be compared with $(144 \pm 8)$ days \cite{modlibra}
which is the measured phase by the DAMA/NaI and DAMA/LIBRA for the low energy peculiar {\it single-hit} rate 
modulation. Thus, the latter one is  
$\gsim$ 5$\sigma$ far from the muon modulation phase measured at LNGS by the large surface apparata
MACRO and LVD.
In conclusion, any possible effect from muons can be safely excluded on the basis of all the given quantitative facts
(and just one of them is enough).

Let us here address in details some recent confusion about the presence of potassium in the detectors 
and its hypotethical role.
As first we remark that the only potassium isotope contributing to the background 
is the radioactive $^{40}$K (natural abundance $1.17 \times 10^{-4}$ and half life $1.248 \times 10^{9}$ yr), 
which is at ppt (10$^{-12}$ g/g) level in the detectors 
\cite{perflibra}; 
it is worth noting that it appears difficult to do better in NaI(Tl), considering e.g. 
the chemical affinity of Na and K. Recently, it was claimed \cite{nat09} 
that some role might be played by $^{40}$K; this possibility 
is evidently discarded by the data, by the published analyses 
and also by simple considerations. 
Let us summarize that:
\begin{itemize}

\item although the peak around 3 keV in the cumulative energy spectrum (see Fig. 1 of ref. \cite{modlibra} and 
      the discussion in ref. \cite{nozzoli})
      can be partially ascribed to $^{40}$K decay, there is not evidence for any 3 keV peak in the S$_m$ distribution 
      (see Fig. 9 of ref. \cite{modlibra}). 
      At the present level of sensitivity the S$_m$ behaviour is compatible within the uncertainties both with a monotonic 
      behaviour and with a {\it kind} of structure, as expected for many Dark Matter candidates and also 
      for WIMPs scenarios; see e.g. the Appendix of ref. \cite{modlibra};

\item $^{40}$K decay cannot give any modulation at all, as well known, unless evoking new exotic physics (see later);

\item no modulation has been observed in other energy regions where $^{40}$K decays also contribute \cite{modlibra};

\item no modulation has been observed in {\it multiple-hits} events (events where more than one detector fires) 
      in the same energy region where DAMA observes the peculiar modulation of the {\it single-hit} events 
      (events where just one detector fires). In fact, $^{40}$K can also give rise to double events in two 
      adjacent detectors when: i) $^{40}$K decays in a detector, $A$, by EC of K shell to the 1461 keV level 
      of $^{40}$Ar; ii) the 1461 keV $\gamma$ escapes from the $A$ detector and hits an adjacent one 
      causing a double coincidence. The 3.2 keV X-rays/Auger electrons from K shell of $^{40}$Ar 
      are fully contained in the $A$ detector with efficiency $\simeq 1$, giving rise 
      in $A$ to a 3.2 keV peak. These double coincidence events, shown in Fig. \ref{fg:40k}, and also possible 
      multi-site events due to Compton scatterings, are 
      {\it multiple-hits} events, and are not modulated, as fore-mentioned (see also ref. \cite{modlibra});

\item no modulation is present just in the double coincidence events (3.2 keV -- 1461 keV) due to $^{40}$K decay.
      In fact, their residuals \footnote{They are calculated averaging -- over the available pairs of adjacent detectors
      in DAMA/LIBRA -- the counting rate of the double coincidence events (3.2 keV -- 1461 keV) 
      due to $^{40}$K decay in every pair, once subtracted
      its mean value.} -- normalized to the error -- as function of the time are shown in Fig. \ref{fg:40k}({\it bottom left})
      throughout the four yearly cycles of DAMA/LIBRA \protect\cite{modlibra}. The two superimposed curves 
      represent the best-fit modulation behaviours $A cos \omega (t-t_0)$ and $A sin \omega (t-t_0)$ 
      with 1 year period ($\omega=2\pi/1 year$) and phase $t_0 = $ June 2nd;
      the best fit values of the modulation amplitudes are: $A=-(0.10\pm0.12)$ and $A=-(0.02\pm0.12)$, respectively.
      These latter values are well compatible with absence of any annual modulation.
      Moreover, the data show a random fluctuation around zero ($\chi^2/dof=1.08$) 
      and their distribution (see Fig. \ref{fg:40k}) is well represented by a gaussian with r.m.s. $\simeq 1$.
      Hence, in conclusion, there is absence of any annual modulation in the double coincidence events due to $^{40}$K decay;

\item fixing the phase of cosine to the perihelion ($\simeq$ Jan 3rd) in the best fit of the data points of
      Fig. \ref{fg:40k}({\it bottom left}), 
      the modulation amplitude is compatible with zero: $A=(0.10\pm0.12)$. Let us note that
      the effect in ref. \cite{fish08} -- where a 0.3\% yearly modulation of nuclear decays of two nuclides 
      has been reported \footnote{Let us remark that in ref. \cite{coop}
      this effect has already been confuted by searching for modifications to the 
      exponential radioactive decay law with the Cassini spacecraft.} 
      with a phase roughly equal to perihelion -- 
      is well below this sensitivity, and this effect is not
      able to account for the DAMA signal because of the well-different phase,
      of the marginal modulation amplitude and of several other arguments given in this list. 

\item the analysis of $^{40}$K double coincidences also rules out at level more than 10 $\sigma$ C.L. 
      the modulation amplitudes expected when assuming that 
      the DAMA effect might be due to the two
      hypothetical cases of: i) $^{40}$K ``exotic'' modulation decay (also see above); 
      ii) spill-out from double to single events and viceversa along the time; on the other hand 
      these exotic hypotetical arguments were already excluded also by several other arguments given in 
      this list.

\item the behaviour of the overall efficiency during the whole data taking period is highly stable. A quantitative investigation
      of the role of the efficiency as possible systematics leads to a cautious upper limit less than 1\% of the 
      modulation amplitude, as reported e.g. in \cite{modlibra};
\begin{figure}[!ht]
\includegraphics[width=0.67\textwidth] {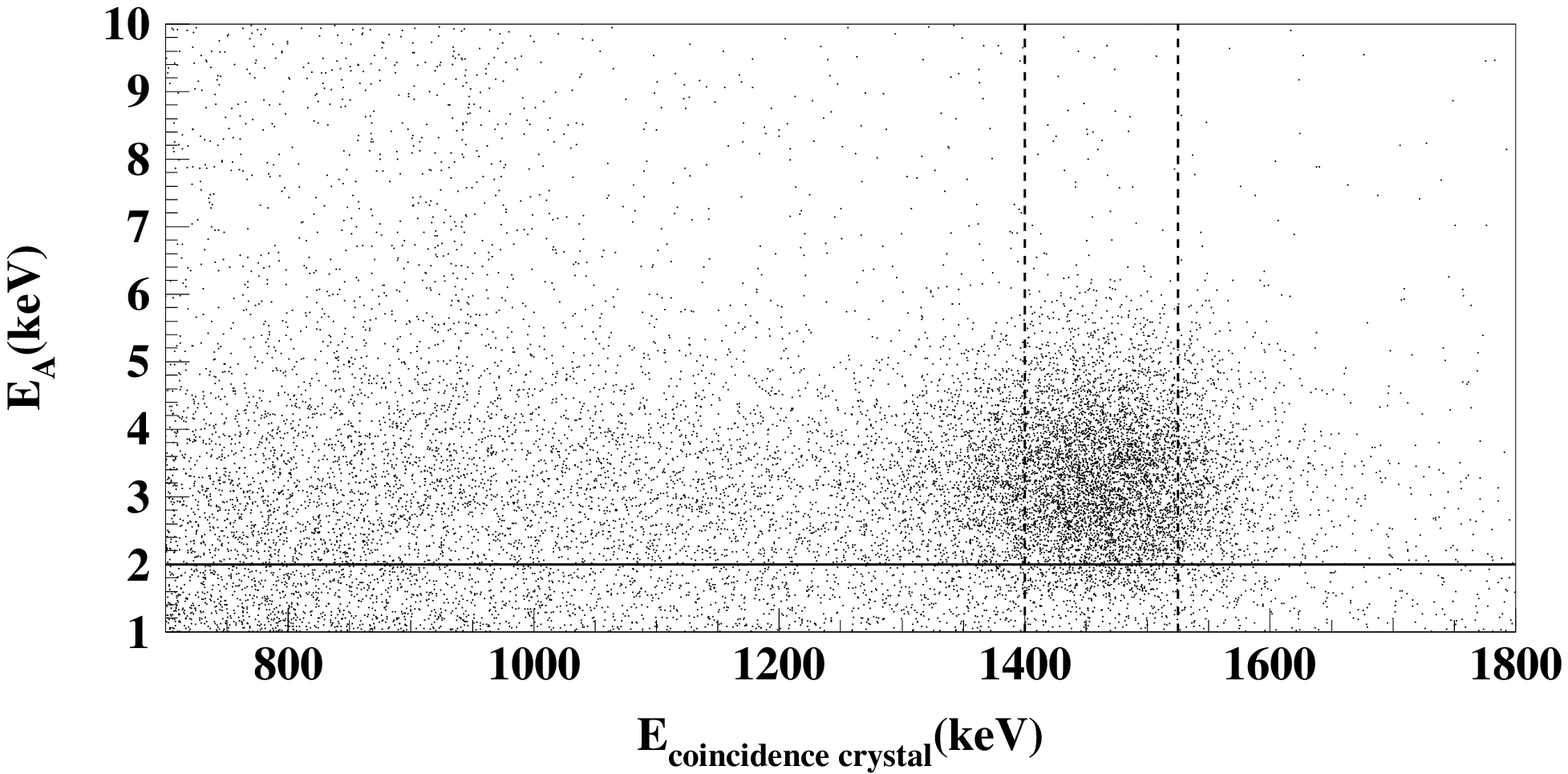}
\end{figure}
\begin{figure}[!ht]
\includegraphics[width=0.37\textwidth] {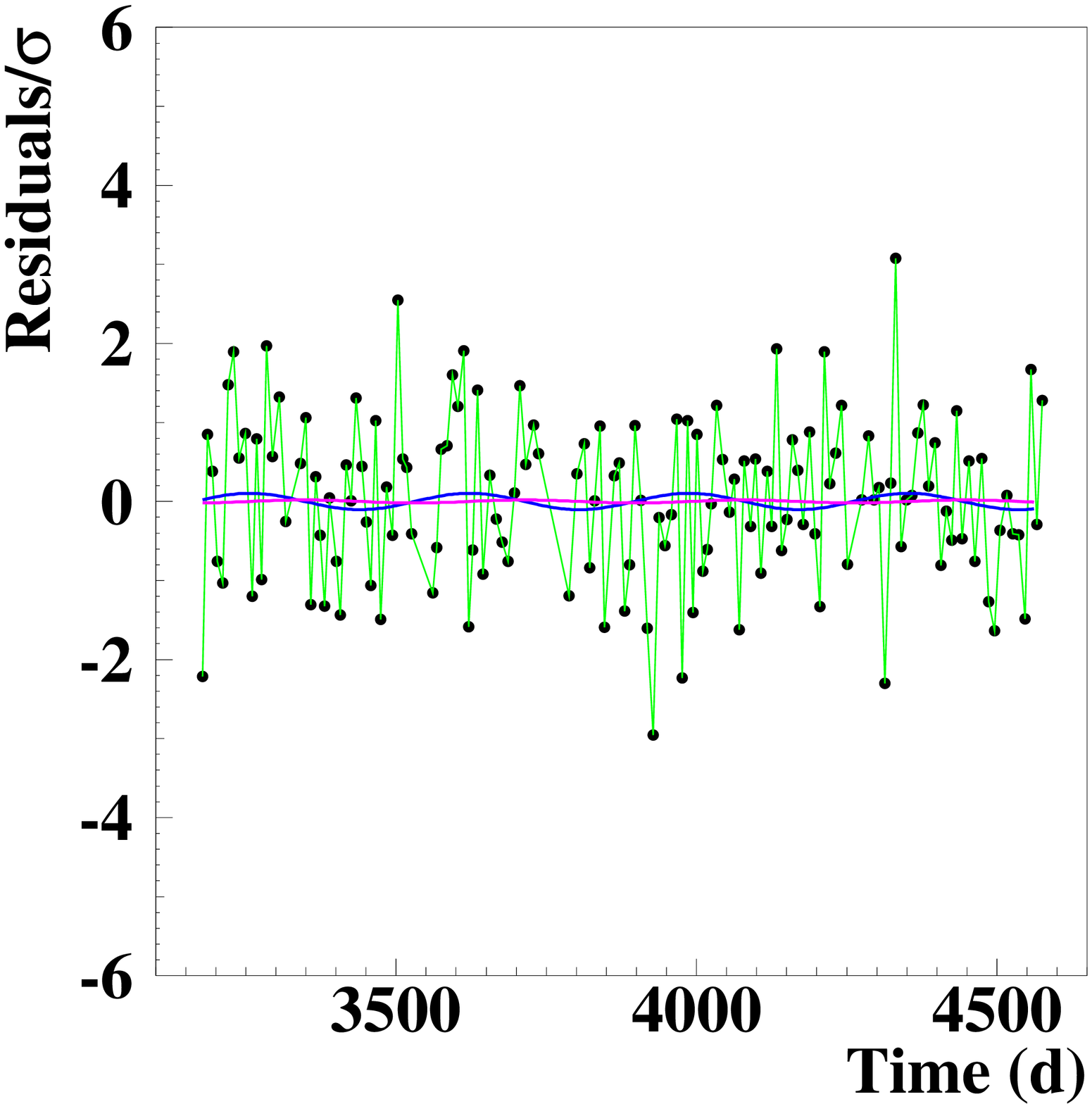}
\includegraphics[width=0.37\textwidth] {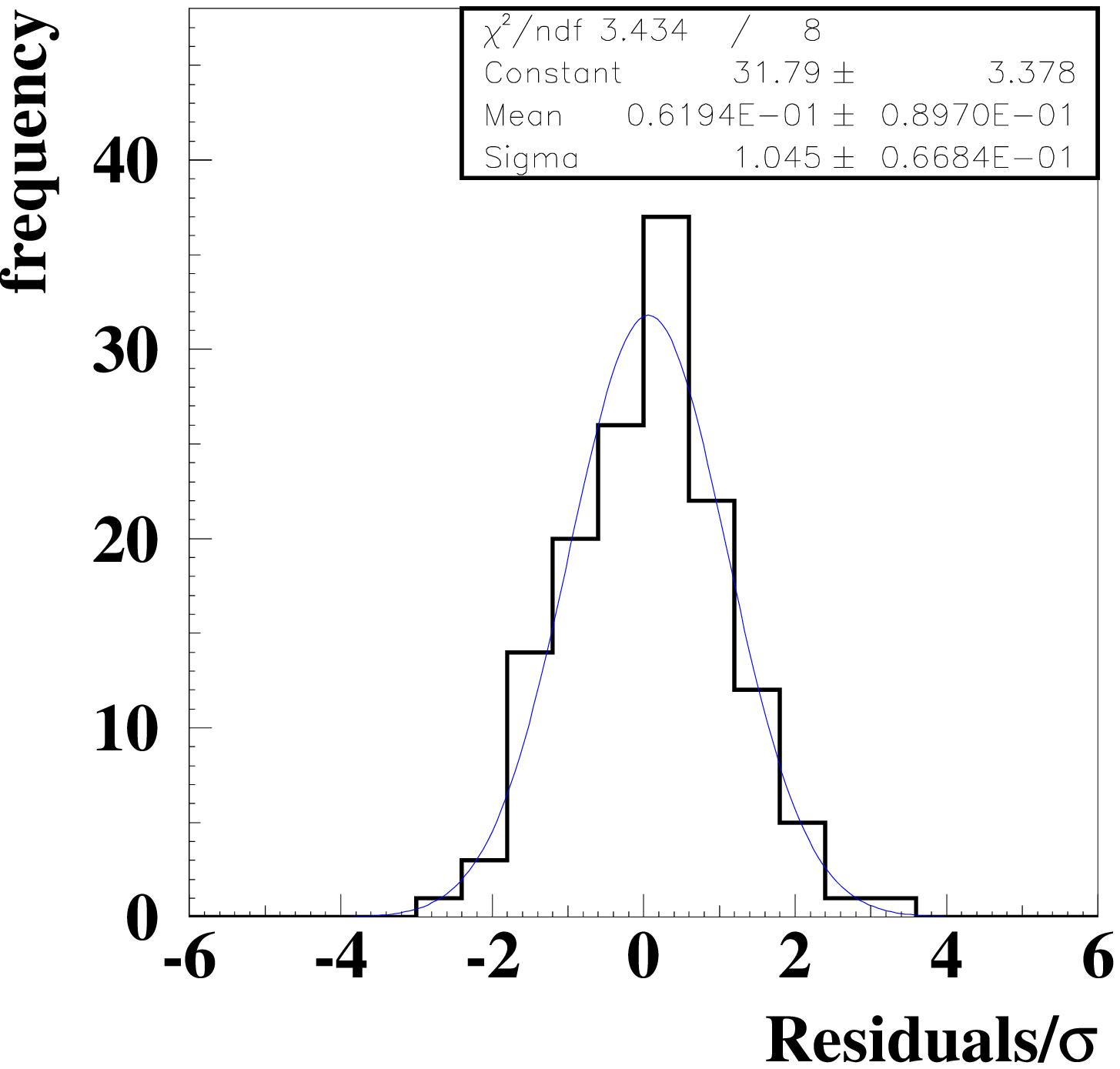}
\vspace{-0.8cm}
\caption{{\it Top:} scatter plot of double coincidence events between a DAMA/LIBRA detector, $A$, (low energy region)
and an adjacent detector (higher energy region).
The threshold of each PMT is at single photoelectron level. For comparison, the software energy threshold
used in the data analyses of the {\it single-hit} events for Dark Matter particle investigation:
2 keV, is shown as continuous line. The double coincidence events due to $^{40}$K decay (3.2 keV -- 1461 keV)
are well identified. For details see ref. \protect\cite{perflibra}.
{\it Bottom left:} residuals -- normalized to the error -- of the double coincidence events (3.2 keV -- 1461 keV) due to
$^{40}$K decay as function of the time
throughout the four yearly cycles of DAMA/LIBRA \protect\cite{modlibra}. The two superimposed curves represent the best-fit
modulation 
behaviours $A cos \omega (t-t_0)$ and $A sin \omega (t-t_0)$ with 1 year period ($\omega=2\pi/1 year$) and 
phase $t_0 = $ June 2nd;
the best fit values of the modulation amplitudes are:
$A=-(0.10\pm0.12)$ and $A=-(0.02\pm0.12)$, respectively. Both values are well compatible with absence of 
any annual
modulation in the double coincidence events due to $^{40}$K decay. Moreover, the data show a random fluctuation around zero
($\chi^2/dof=1.08$) and their distribution ({\it bottom right panel}) is well represented by a gaussian with r.m.s. $\simeq 1$.}
\label{fg:40k}
\vspace{-0.8cm}
\end{figure}
\item the annual modulation signal -- observed by DAMA -- is present both in the outer and in the inner detectors \cite{modlibra}. 
      Hence, there is no dependence on the veto capability, that is different -- by geometrical reasons -- 
      among the outer and the inner detectors.
      In particular, if the $^{40}$K decay would hypothetically play some role in the annual modulation of the low-energy 
      {\it single-hit} events, the effect would be larger in the outer detectors where the 1461 keV $\gamma$'s
      accompanying the 3.2 keV X-rays/Auger electrons have lower probability to be detected by closer detectors; 

\item the annual modulation signal -- observed by DAMA -- is equally distributed over all the detectors \cite{modlibra};
      see also the previous considerations.

\end{itemize}

Thus, for all the above reasons (and just one of them is enough), no role can be played by $^{40}$K.

In conclusion, DAMA/LIBRA has confirmed the presence of an annual modulation satisfying all the 
requirements of the DM annual 
modulation signature, as previously pointed out by DAMA/NaI; in particular, the evidence 
for the presence of DM particles in the galactic halo is cumulatively supported at 8.2 $\sigma$ C.L..

It is worth noting that no other experiment exists, whose result can be directly compared in a 
model-independent way with those by DAMA/NaI and DAMA/LIBRA.
Moreover, concerning those activities claiming for some model dependent exclusion under some set of 
largely arbitrary 
assumptions (see for example discussions in \cite{RNC,paperliq,benoit}),
some important critical points exist in some of their experimental aspects 
(energy threshold, energy scale, multiple selection procedures, 
disuniformity of the detectors response, absence of suitable periodical 
calibrations in the same running conditions and in the claimed low energy region, 
stabilities, etc.). 

Finally, as regards the indirect detection searches, let us note 
that no direct model-independent comparison
can be performed between the results obtained in direct and indirect activities,
since it does not exist a biunivocal correspondence between the observables in the
two kinds of experiments. 
Anyhow, if possible excesses in the positron to electron flux ratio 
and in the $\gamma$ rays flux with respect to an assumed simulation of the hypothesized background contribution,
which is expected from standard sources,
might be interpreted in terms of Dark Matter (but huge and still unjustified boost factor and new interaction
types are required), 
this would also be not in conflict with the effect observed by DAMA experiments.

\vspace{-0.3cm}
\section{UPGRADES AND PROSPECTS}

A first upgrade of the DAMA/LIBRA set-up was performed in September 2008. 
One detector was recovered by replacing a broken PMT and a new optimization of some PMTs and HVs was done.
The transient digitizers were replaced with new ones, having better performances and a new DAQ with optical read-out was installed;
since October 2008 DAMA/LIBRA is again in operation. Data of
two further annual cycles are at hand.

Considering the relevance to lower the software energy threshold of the experiment,
in order to improve the performance and the sensitivity of the experiment 
and to allow also deeper corollary information on the nature of the 
DM candidate particle(s) and on the various related 
astrophysical, nuclear and particle Physics scenarios, 
the replacement of all the PMTs with new ones 
with higher quantum efficiency has been planned and work is in progress.
DAMA/LIBRA will also study several other rare
processes as done by the former DAMA/NaI apparatus in the past \cite{allRare}
and by itself so far \cite{papep}.

\end{document}